Authors: Kingshuk Chatterjee[1], Kumar Sankar Ray(corresponding author)[2]

Affiliations:
[1]Electronics and Communication Sciences Unit, Indian Statistical Institute, Kolkata-108

[2]Professor, Electronics and Communication Sciences Unit, Indian Statistical Institute, Kolkata-108

Address: Electronics and Communication Sciences Unit, Indian Statistical Institute, Kolkata-108.

Telephone Number: +918981074174

Fax Number:033-25776680

Email:ksray@isical.ac.in

# Restricted deterministic Watson-Crick automata

*Kingshuk Chatterjee,[1] Kumar Sankar Ray[2]*

*Electronics and Communication Science unit, ISI, Kolkata.*

[1]kingshukchaterjee@gmail.com [2]ksray@isical.ac.in

*Abstract: In this paper, we introduce a new model of deterministic Watson-Crick automaton namely restricted deterministic Watson-Crick automaton which is a deterministic Watson-Crick automaton where the complementarity string in the lower strand is restricted to a language L. We examine the computational power of the restricted model with respect to L being in different language classes such as regular, unary regular, finite, context free and context sensitive. We also show that computational power restricted deterministic Watson-Crick automata with L in regular languages is same as that of deterministic Watson-Crick automata and that the set of all languages accepted by restricted deterministic Watson-Crick automata with L in unary regular languages is a proper subset of context free languages.*

## I. INTRODUCTION

Watson-Crick automata [1] are finite automata having two independent heads working on double strands where the characters on the corresponding positions of the two strands are connected by a complementarity relation similar to the Watson-Crick complementarity relation. The movement of the heads although independent of each other is controlled by a single state. Non-deterministic Watson-Crick automata have been explored in [2]. Deterministic Watson-Crick automata and its variants have been explicitly handled in [3]. Parallel Communicating Watson-Crick automata were introduced in [4] and further investigated in [5]. Equivalence of subclasses of two-way Watson-Crick automata is discussed in [6]. A survey of Watson-Crick automata can be found in [7]. Research work regarding state complexity of Watson-Crick automata is reported in [8] and [9].

In this paper, we introduce a new model of deterministic Watson-Crick automata namely restricted deterministic Watson-Crick automata which are deterministic Watson-Crick automata where the complementarity string in the lower strand is restricted to a language  L. Thus this restricted automaton can accept a subset of only those strings whose complementarity string is in L. We study the effect of the restriction on the lower strand on the model. We see how the different restriction languages effect the computational power of the restricted deterministic Watson-Crick automata. We show that restricted deterministic Watson-Crick automata with lower strand restricted to regular languages has the same computational power as deterministic Watson-Crick automata. We further show that set of all languages accepted by restricted deterministic Watson-Crick automata with lower strand restricted to unary regular languages is a proper subset of the context free languages. We also examine the computational power of restricted automata where the lower strand is restricted to context free and context sensitive languages.

## II. BASIC TERMINOLOGY

The symbol V denotes a finite alphabet. The set of all finite words over V is denoted by $V^*$, which includes the empty word $\lambda$. The symbol $V^+ = V^* - \{\lambda\}$ denotes the set of all non-empty words over the alphabet V. For $w \in V^*$, the length of w is denoted by |w|. Let $u \in V^*$ and $v \in V^*$ be two words and if there is some word $x \in V^*$, such that v=ux, then u is a prefix of v, denoted by $u \leq v$. Two words, u and v are prefix comparable denoted by $u \sim_p v$ if u is a prefix of v or vice versa.

A Watson-Crick automaton is a 6-tuple of the form M=(V,$\rho$,Q,$q_0$,F,$\delta$) where V is an  alphabet set, set of states is denoted by Q, $\rho \subseteq V \times V$ is the complementarity relation similar to Watson-Crick complementarity relation, $q_0$ is the initial state and F⊆Q is the set of final states. The function $\delta$ contains a finite number of transition rules of the form $q\binom{w_1}{w_2} \rightarrow q'$, which denotes that the machine in state q parses $w_1$ in upper strand and $w_2$ in lower strand and goes to state q' where $w_1, w_2 \in V^*$. The symbol $\begin{bmatrix} w_1 \\ w_2 \end{bmatrix}$ is different from $\binom{w_1}{w_2}$. While $\binom{w_1}{w_2}$ is just a pair of strings written in that form instead of $(w_1, w_2)$, the symbol $\begin{bmatrix} w_1 \\ w_2 \end{bmatrix}$ denotes that the two strands are of same length i.e. $|w_1|=|w_2|$ and the corresponding symbols in two strands are complementary in the sense given by the relation $\rho$.  The symbol $\begin{bmatrix} V \\ V \end{bmatrix}_\rho = \{ \begin{bmatrix} a \\ b \end{bmatrix} \mid a, b \in V, (a, b) \in \rho \}$ and $WK_\rho(V) = \begin{bmatrix} V \\ V \end{bmatrix}_\rho^*$ denotes the Watson-Crick domain associated with V and $\rho$.

A transition in a Watson-Crick finite automaton can be defined as follows:

For $\binom{x_1}{x_2}, \binom{u_1}{u_2}, \binom{w_1}{w_2} \in \binom{V^*}{V^*}$ such that $\begin{bmatrix} x_1 u_1 w_1 \\ x_2 u_2 w_2 \end{bmatrix} \in WK_\rho(V)$ and $q, q' \in$ Q, $\binom{x_1}{x_2} q \binom{u_1}{u_2} \binom{w_1}{w_2} \Rightarrow \binom{x_1}{x_2} \binom{u_1}{u_2} q' \binom{w_1}{w_2}$ iff there is transition rule $q\binom{u_1}{u_2} \rightarrow q'$ in $\delta$ and $\overset{*}{\Rightarrow}$ denotes the transitive and reflexive closure of $\Rightarrow$. The language accepted by a Watson-

Crick automaton M is L(M)={$w_1 \in V^*$|$q_0 \begin{bmatrix} w_1 \\ w_2 \end{bmatrix} \stackrel{*}{\Rightarrow} q \begin{bmatrix} \lambda \\ \lambda \end{bmatrix}$, with q ∈ F, $w_2 \in V^*$, $\begin{bmatrix} w_1 \\ w_2 \end{bmatrix} \in WK_\rho(V)$}.

Depending on the type of states and transition rules there are four types or subclasses of Watson-Crick automata. A Watson-Crick automaton M=(V,ρ,Q,$q_0$,F, δ) is

1) stateless( NWK ): If it has only one state, i.e. Q=F={ $q_0$ };
2) all-final( FWK ): If all the states are final, i.e. Q=F;
3) simple( SWK ): If at each step the automaton reads either from the upper strand or from the lower strand, i.e. for any transition rule q$\begin{pmatrix} w_1 \\ w_2 \end{pmatrix}$→q', either $w_1$= λ or $w_2$= λ;
4) 1-limlited( 1-limited WK ): If for any transition rule q$\begin{pmatrix} w_1 \\ w_2 \end{pmatrix}$→q', we have |$w_1 w_2$|=1.

## III. DETERMINISTIC WATSON-CRICK AUTOMATA AND THEIR SUBCLASSES

The notion of determinism in Watson-Crick automata and a discussion on its complexity were first considered in [3]. In [3] different notions of determinism were suggested as follows:

1) weakly deterministic Watson-Crick automata(WDWK): Watson-Crick automaton is weakly deterministic if in every configuration that can occur in some computation of the automaton, there is a unique possibility to continue the computation, i.e. at every step of the automaton there is at most one way to carry on the computation.
2) deterministic Watson-Crick automata(DWK): deterministic Watson-Crick automaton is Watson-Crick automaton for which if there are two transition rules of the form q$\begin{pmatrix} u \\ v \end{pmatrix}$→q' and q$\begin{pmatrix} u' \\ v' \end{pmatrix}$→q" then u$\not\prec_p$u' or v$\not\prec_p$v'.
3) strongly deterministic Watson-Crick automata(SDWK): strongly deterministic Watson-Crick automaton is a deterministic Watson-Crick automaton where the Watson-Crick complementarity relation is injective.

Similar to non-deterministic Watson-Crick automata, deterministic Watson-Crick automata can be stateless (NDWK), all final (FDWK), simple (SiDWK) and 1-limited (1-limited DWK).

## IV. RESTRICTED DETERMINISTIC WATSON-CRICK AUTOMATA

A restricted deterministic Watson-Crick automaton is derived from deterministic Watson-Crick automaton by including a additional restriction on its lower tape in the definition of the model. Informally, the restriction is imposed as follows:

Along with the model of a deterministic Watson-Crick automaton we also describe a Language L in a restricted deterministic Watson-Crick automaton. Only those strings are allowed as input to the automaton whose complementarity string (obtained by applying the complementarity relation ρ on the string in the upper strand) is in L.

Formally, a restricted deterministic Watson-Crick automaton is a 7-tuple of the form M=(V,ρ,Q,$q_0$,F,δ,L) where V is an alphabet set, set of states is denoted by Q, ρ ⊆ V×V is the complementarity relation similar to Watson-Crick complementarity relation, $q_0$ is the initial state and F⊆Q is the set of final states. The function δ contains a finite number of transition rules of the form q$\begin{pmatrix} w_1 \\ w_2 \end{pmatrix}$→q', in δ, if there are two transition rules of the form q$\begin{pmatrix} u \\ v \end{pmatrix}$→q' and q$\begin{pmatrix} u' \\ v' \end{pmatrix}$→q" then u$\not\prec_p$u' or v$\not\prec_p$v'. The language to which the strings in the lower strand of the input to the automaton must belong to is denoted by L. That is if no complementarity string of the string w belong to L. Then w is not allowed as input to the automaton and is out rightly rejected without executing the automaton on w.

A transition in a restricted Watson-Crick finite automaton can be defined as follows:

For $\begin{pmatrix} x_1 \\ x_2 \end{pmatrix}, \begin{pmatrix} u_1 \\ u_2 \end{pmatrix}, \begin{pmatrix} w_1 \\ w_2 \end{pmatrix} \in \begin{pmatrix} V^* \\ V^* \end{pmatrix}$ such that $\begin{bmatrix} x_1 u_1 w_1 \\ x_2 u_2 w_2 \end{bmatrix} \in WK_\rho(V)$, $x_2 u_2 w_2 \in L$ and $q, q' \in Q$, $\begin{pmatrix} x_1 \\ x_2 \end{pmatrix} q \begin{pmatrix} u_1 \\ u_2 \end{pmatrix} \begin{pmatrix} w_1 \\ w_2 \end{pmatrix} \Rightarrow \begin{pmatrix} x_1 \\ x_2 \end{pmatrix} \begin{pmatrix} u_1 \\ u_2 \end{pmatrix} q' \begin{pmatrix} w_1 \\ w_2 \end{pmatrix}$ iff there is transition rule q$\begin{pmatrix} u_1 \\ u_2 \end{pmatrix}$→q' in δ and $\stackrel{*}{\Rightarrow}$ denotes the transitive and reflexive closure of ⇒.

The language accepted by a restricted Watson-Crick automaton M is L(M)={$w_1 \in V^*$|$q_0 \begin{bmatrix} w_1 \\ w_2 \end{bmatrix} \stackrel{*}{\Rightarrow} q \begin{bmatrix} \lambda \\ \lambda \end{bmatrix}$, with q ∈ F, $w_2 \in L$, $\begin{bmatrix} w_1 \\ w_2 \end{bmatrix} \in WK_\rho(V)$}.

## V. COMPUTATIONAL COMPLEXITY OF RESTRICTED DETERMINISTIC WATSON-CRICK AUTOMATA

In this section, we discuss the computational complexity of restricted deterministic Watson-Crick automata.

**Theorem 1:** For every deterministic Watson-Crick automaton there exists a restricted deterministic Watson-Crick automaton which accepts the same language and has L=$V^*$.

**Proof:** From the definition of deterministic Watson-Crick automaton and restricted deterministic Watson-Crick automaton, we see that the only difference between the two automaton is in their accepting condition.

The accepting condition for deterministic Watson-Crick automaton is L'(M)={$w_1 \in V^*$|$q_0 \begin{bmatrix} w_1 \\ w_2 \end{bmatrix} \overset{*}{\Rightarrow} q \begin{bmatrix} \lambda \\ \lambda \end{bmatrix}$}, with q ∈ F, $w_2 \in V^*$, $\begin{bmatrix} w_1 \\ w_2 \end{bmatrix} \in WK_\rho(V)$} whereas the accepting condition for restricted deterministic Watson-Crick automaton is L'(M)={$w_1 \in V^*$|$q_0 \begin{bmatrix} w_1 \\ w_2 \end{bmatrix} \overset{*}{\Rightarrow} q \begin{bmatrix} \lambda \\ \lambda \end{bmatrix}$}, with q ∈ F, $w_2 \in L$, $\begin{bmatrix} w_1 \\ w_2 \end{bmatrix} \in WK_\rho(V)$}. In the accepting condition for deterministic Watson-Crick automaton $w_2 \in V^*$ but in the accepting condition of restricted deterministic Watson-Crick automaton $w_2 \in L$ where L is the language to which the strings in the lower strand of the restricted model must belong.

Now if $L=V^*$ in the restricted deterministic Watson-Crick automaton then the accepting condition for both the models become the same. Thus, for every deterministic Watson-Crick automaton there exists a restricted deterministic Watson-Crick automaton which has all its parameter same as the deterministic Watson-Crick automaton and has $L=V^*$ and accepts the same language.

**Lemma 1:** For every restricted deterministic Watson-Crick automaton there exists a 1-limited restricted deterministic Watson-Crick automaton which accepts the same language and has the same restriction language L.

**Proof:** The proof is exactly same as the proof by Czeizler et.al. in [3].

**Theorem 2:** For every restricted 1-limited deterministic Watson-Crick automaton where the restriction language is regular there exists a deterministic Watson-Crick automaton which accepts the same language as the restricted 1-limited deterministic Watson-Crick automaton.

**Proof:** As L is regular we can obtain a deterministic finite automaton M'=(Q', V', $q_0$', F', δ') which accepts L. The deterministic Watson-Crick automaton M"=(V,ρ",Q",$q_0$",F", δ") is constructed from the restricted 1-limited deterministic Watson-Crick automaton M=(V,ρ,Q,$q_0$,F,δ,L) in the following manner:

The finite alphabet V is same in both the automaton, the set of states Q"=Q×Q', the start state $q_0$"=($q_0$,$q_0$'), the set of final states F"={($q_i$,$q_j$)| $q_i \in F$ and $q_j \in F'$} and the complementarity relation ρ is same in both the automaton.

The transitions in δ" are obtained from δ and δ' as follows:

As M is 1-limited therefore the transitions in δ are of two types.

1) $q_i \binom{x}{\lambda} \rightarrow q_j$ and $x \in V$ and $q_i, q_j \in Q$,

2) $q_i \binom{\lambda}{x} \rightarrow q_j$ and $x \in V$ and $q_i, q_j \in Q$.

For each transition of type 1 in δ the following transitions are introduced in δ".

$(q_i,q_k) \binom{x}{\lambda} \rightarrow (q_j,q_k)$ and $x \in V$, $q_i, q_j \in Q$ for all $q_k \in Q'$.

For each transition of type 2 in δ the following transition is introduced in δ".

$(q_i,q_k) \binom{\lambda}{x} \rightarrow (q_j,q_l)$ and $x \in V$, $q_i, q_j \in Q$ and $q_l \in Q'$ and there is a transition $δ(q_k,x)=q_l$ in M' for all $q_k \in Q'$,

This ensures that every time M" reads its lower strand to simulate M it also simulates the deterministic finite automaton M'.

Now if a string w is accepted by the restricted 1-limited deterministic Watson-Crick automaton M then the complementarity string of w must be in L and the restricted 1-limited deterministic Watson-Crick automaton must be in its final state after completely parsing w and its complementarity string in both its strands. Thus the deterministic Watson-Crick automaton M" simulating M and M' must also have its state as ($q_i,q_k$) after completely parsing both its strands where $q_i \in F$ and $q_j \in F'$ as a result ($q_i,q_k$) is in F" thus M" accepts w.

If a string w is not accepted by the restricted 1-limited deterministic Watson-Crick automaton M. Then w is rejected due to occurrence of one of the following events.

1) The complementarity string of w is not in L, then the deterministic Watson-Crick automaton M" simulating M at the end of parsing the input will never be in a state ($q_i,q_k$) where $q_k \in F'$. Therefore M" will not be in its final state and thus M" will also reject w.

2) The restricted 1-limited deterministic Watson-Crick automaton M halts before w is completely parsed. As the deterministic Watson-Crick automaton M" simulates M it will also halt due to lack of transition that can be applied without completely traversing w thus M" will also reject w.

3) If the restricted 1-limited deterministic Watson-Crick automaton M after completely parsing both its strand is in a non-final state. As the deterministic Watson-Crick automaton M" simulates M at the end of parsing its input strands completely will never be in a state $(q_i,q_k)$ where $q_i \in F$. Thus M'' will not be in its final state and so M'' will also reject w.

So we can say that the deterministic Watson-Crick automaton M'' constructed from M and M' accepts the same language as the restricted 1-limited deterministic Watson-Crick automaton M.

**Corollary 1:** For every restricted deterministic Watson-Crick automaton where the restriction language is regular there exists a deterministic Watson-Crick automaton which accepts the same language as the restricted deterministic Watson-Crick automaton.

**Proof:** The proof follows from Lemma 1 and Theorem 2.

**Lemma 2:** The set of all languages accepted by deterministic Watson-Crick automata is a subset of the set of all languages accepted by restricted deterministic Watson-Crick automata where L is regular.

**Proof:** As $L=V^*$ is regular and from Theorem 1 we know that for every deterministic Watson-Crick automaton there exists a restricted deterministic Watson-Crick automaton with $L=V^*$ which accepts the same language. Thus the set of all languages accepted by restricted deterministic Watson-Crick automata where L is regular contains the set of all languages accepted by deterministic Watson-Crick automata. Thus the set of all languages accepted by deterministic Watson-Crick automata is the subset of set of all languages accepted by restricted deterministic Watson-Crick automata where L is regular.

**Lemma 3:** The set of all languages accepted by restricted deterministic Watson-Crick automata where L is regular is a subset of the set of all languages accepted by deterministic Watson-Crick automata.

**Proof:** From Corollary 1, we know that for every restricted deterministic Watson-Crick automaton where L is regular there exists a deterministic Watson-Crick automaton which accepts the same language. Thus the set of all languages accepted by restricted deterministic Watson-Crick automata where L is regular is the subset of the set of all languages accepted by deterministic Watson-Crick automata.

**Theorem 3:** The computation power of deterministic Watson-Crick automata and restricted deterministic Watson-Crick automata where L is regular are equal.

**Proof:** From Lemma 2, we know that the set of all languages accepted by deterministic Watson-Crick automata is a subset of the set of all languages accepted by restricted deterministic Watson-Crick automata where L is regular and Lemma 3 states that the set of all languages accepted by restricted deterministic Watson-Crick automata where L is regular is a subset of the set of all languages accepted by deterministic Watson-Crick automata. Thus, we can conclude, the set of all languages accepted by deterministic Watson-Crick automata and the set of all languages accepted by restricted deterministic Watson-Crick automata where L is regular are equal. So the computational power of deterministic Watson-Crick automata and restricted deterministic Watson-Crick automata where L is regular are equal.

**Corollary 2:** Restricted deterministic Watson-Crick automata where L is regular can accept some context free and context sensitive languages.

**Proof:** The proof follows from Theorem 3. As Theorem 3 states that the set of all languages accepted by restricted deterministic Watson-Crick automata with regular L is equal to the set of all languages accepted by deterministic Watson-Crick automata and from [3] we know that deterministic Watson-Crick automata can accept some context free and context sensitive languages. Thus we can conclude, restricted deterministic Watson-Crick automata where L is regular can accept some context free and context sensitive languages.

**Theorem 4:** For every regular language $L' \subseteq V^*$ there exists a restricted deterministic Watson-Crick automaton where $L=a^*$ which accepts L'.

**Proof:** As L is regular we can obtain a deterministic finite automaton $M'=(Q', V', q_0', F', \delta')$ which accepts L. The restricted deterministic Watson-Crick automaton $M=(V,\rho,Q,q_0,F,\delta,L)$ with $L=a^*$ is constructed from M as follows:

The finite alphabet $V=V' \cup \{a\}$, the set of states $Q=Q'$, the start state $q_0=q_0'$, the set of final states $F=F'$ and the complementarity relation $\rho$ is as follows:

$\rho(x)=a$ for all $x \in V'$.

The transitions in $\delta$ are obtained from $\delta'$ in the following manner:

For every transition $\delta(q_i,x)=q_j$ in M' where $q_i, q_j \in Q'$ and $x \in V'$ the transition $q_i \binom{x}{a} \rightarrow q_j$ where $q_i, q_j \in Q'$ and $x \in V'$ is introduced in $\delta'$. The upper strand of the restricted deterministic Watson-Crick automaton M simulates M' whereas the lower strand just reads a.

Now for a string w in L', the deterministic finite automaton M' will go to its final state after consuming w. As, M simulates M' it will also go its final state after consuming its input. Thus M will also accept w.

If w is not in L', the deterministic finite automaton M' will not go to its final state after consuming w. As, M simulates M' it will also not go its final state after consuming its input. Thus M will also reject w.

**Example 1:** The following restricted deterministic Watson-Crick automaton M is an example of a restricted deterministic Watson-Crick automaton whose lower strand is restricted to $L=a^+$ which accepts the language $L'=\{a^n b^n \mid n>1\}$ where L' is non-regular.

$M=(\{a,b\}, \{q_0,q_f\}, q_0, \{q_f\}, \delta, \rho, L)$, $L=a^+$ where the complementarity relation $\rho$ is as follows: $\rho(a)=a$, $\rho(b)=a$ and the transitions in $\delta$ are as follows:

$q_0 \binom{a}{\lambda} \to q_0$, $q_0 \binom{b}{aa} \to q_f$, $q_f \binom{b}{aa} \to q_f$.

**Theorem 5:** There exists a restricted deterministic Watson-Crick automaton with its lower strand restricted to $a^+$ which accepts a non-regular language.

**Proof:** The proof follows from Example 1.

**Theorem 6:** The set of all languages accepted by restricted deterministic Watson-Crick automata whose lower strand is restricted to unary regular languages is a proper superset of regular language.

**Proof:** From Theorem 4, we see that for every regular language L' we can obtain a restricted deterministic Watson-Crick automaton with $L=a^*$ which accepts L'. As $L=a^*$ is a unary regular language therefore regular languages is a subset of the set of all languages accepted by restricted deterministic Watson-Crick automata where L is a unary regular language. From Example 1, we see that there is a restricted deterministic Watson-Crick automaton with L in unary regular language which accepts a non-regular language. Hence, the set of all languages accepted by restricted deterministic Watson-Crick automata whose lower strand is restricted to unary regular languages is a proper superset of regular language.

**Theorem 7:** For every restricted 1-limited deterministic Watson-Crick automaton where L is a unary regular language there exists a non-deterministic pushdown automaton which accepts by empty stack that accepts the same language as the restricted deterministic Watson-Crick automaton.

**Proof:** Let $M=(V,\rho,Q,q_0,F,\delta,L)$ be the restricted 1-limited deterministic Watson-Crick automaton with L in unary regular languages. Without loss of generality, let us assume $a \in V$ and $b \notin V$ and the unary regular language L is defined on 'a'. As the strings in the lower strand is restricted to 'a'. Thus, the order or arrangement of the alphabet becomes immaterial and only the number of a's present in the lower strand becomes important. Hence, for a restricted 1-limited deterministic Watson-Crick automaton where L is in unary regular languages defined on 'a'. Reading the lower strand from beginning or end is the same. We only have to make sure that in simulating M the number of a's read is equal to |w| where w is the input to the restricted 1-limited deterministic Watson-Crick automaton and |w| is the length of |w|. Thus, the reading of the lower strand can be simulated by a stack. moreover as the lower strand is restricted to a unary regular language L we can obtain a deterministic finite automaton $M'=(Q', \{a\}, q_0', F', \delta')$ which accepts L.

Now, we aim to simulate the moves of M using a non-deterministic pushdown automaton $NP=(Q'', V, \Gamma, q_0'', \$, \delta'')$ which accepts by empty stack. The non-deterministic pushdown automaton NP is constructed from M and M' in the following manner:

The finite alphabet V is same in both the automaton, the set of states $Q''=Q \times Q'$, the start state $q_0''=(q_0,q_0')$, the set of stack symbols $\Gamma=\{a,b,\$\}$ where $\$$ is the initial stack symbol. The transitions in $\delta''$ is obtained from $\delta$ and $\delta'$ as follows:

As M is 1-limited therefore the transitions in $\delta$ are of two types.

1) $q_i \binom{x}{\lambda} \to q_j$ and $x \in V$ and $q_i, q_j \in Q$,

2) $q_i \binom{\lambda}{x} \to q_j$ and $x \in V$ and $q_i, q_j \in Q$.

The number of a's in the stack represents by how much the upper head is ahead of the lower head in the restricted 1-limited deterministic Watson-Crick automaton M which NP is simulating at any given stage. The number of b's in the stack represent by how much the lower head is ahead of the upper head in the restricted 1-limited deterministic Watson-Crick automaton M which NP is simulating at any given stage. The initial stack symbol at the top of the stack represents the fact that both heads of M are in the same position.

For each transition of type 1 in $\delta$ the following transitions are introduced in $\delta''$.

$\delta''((q_i,q_k),x,a)=((q_j,q_k), aa)$, $\delta''((q_i,q_k),x,\$)=((q_j,q_k), \$a)$ and $\delta''((q_i,q_k),x,b)=((q_j,q_k), \lambda)$ where $q_i, q_j \in Q$, for all $q_k \in Q'$ and $x \in V$.

In the first transition, when simulating the reading of the upper head the input to NP is read and if the stack has 'a' at its top then the upper head of the restricted 1-limited deterministic Watson-Crick automaton M which NP is simulating is ahead of the lower head at that stage thus reading of the upper head will further increment the distance thus while representing the consumption of the upper head character by NP an 'a' is introduced in the stack.

In the second transition, when simulating the reading of the upper head the input to NP is read and if the stack has $ at its top then the upper head of the restricted 1-limited deterministic Watson-Crick automaton M which NP is simulating is at the same position as the lower head at that stage thus reading of the upper head will move the upper head by 1 thus while representing the consumption of the upper head character by NP an 'a' is introduced in the stack.

In the third transition, when simulating the reading of the upper head the input to NP is read and if the stack has 'b' at its top then the upper head of the restricted 1-limited deterministic Watson-Crick automaton M which NP is simulating is behind the lower ahead at that stage thus reading of the upper head will decrement the distance thus while representing the consumption of the upper head character by NP 'b' is popped from the stack.

For transitions of the form $q_i\binom{\lambda}{x} \to q_j$ and $x \in V$ and $q_i, q_j \in Q$ in M only those transactions having x=a is valid as the lower strand is restricted to unary language defined on a. Thus a transaction $q_i\binom{\lambda}{x} \to q_j$ and $x \in V$ and $q_i, q_j \in Q$ and $x \neq a$ can never be applied in M.

Thus the transitions of type 2 reduces to $q_i\binom{\lambda}{a} \to q_j$ where $q_i, q_j \in Q$.
For each such transition in δ the following transitions are introduced in δ".

δ"$((q_i,q_k),\lambda,a)=((q_j,q_l), \lambda)$, δ"$((q_i,q_k),\lambda,\$)=((q_j,q_k), \$b)$ and δ"$((q_i,q_k),\lambda,b)=((q_j,q_k), bb)$ where $q_i, q_j \in Q$, for all $q_k \in Q$, $q_l \in Q'$, $x \in V$ and there is a transition $\delta(q_k,x)=q_l$ in M'.

In the first transition, when simulating the reading of the lower head the operation is done on the stack of NP and if the stack has 'a' at its top then the lower head of the restricted 1-limited deterministic Watson-Crick automaton M which NP is simulating is behind the lower head at that stage thus simulation of the reading of the lower head should decrement the distance thus while representing the consumption of the lower head character by NP an 'a' is popped from the stack.

In the second transition, when simulating the reading of the lower head the operation is done on the stack of NP and if the stack has $ at its top then the lower head of the restricted 1-limited deterministic Watson-Crick automaton M which NP is simulating is at the same position as the upper head at that stage thus reading of the lower head will move the lower head ahead by 1 thus while representing the consumption of the lower head character by NP a 'b' is introduced in the stack.

In the third transition, when simulating the reading of the lower head the operation is done on the stack of NP and if the stack has 'b' at its top then the lower head of the restricted 1-limited deterministic Watson-Crick automaton M which NP is simulating is ahead of the upper ahead at that stage thus reading of the lower head will further increase the distance thus while representing the consumption of the lower head character by NP 'b' is introduced in the stack.

Moreover in simulating the lower head of the restricted 1-limited deterministic Watson-Crick automaton M, the non-deterministic pushdown automaton NP also simulates the deterministic finite automaton M'.

In addition to these transitions introduced in δ" with respect to transitions in δ and δ', another set of transitions are introduced in δ".

δ"$((q_f,q_k),\lambda,\$)=((q_f,q_k), \lambda)$ where $q_f \in F$, and $q_k \in F'$.

These transitions ensures that the non-deterministic pushdown automaton NP in simulating the restricted 1-limited deterministic Watson-Crick automaton M reaches the empty stack condition after consuming its input when M reaches its final state after both its head reach the end of the input.

Now if a string w is accepted by the restricted 1-limited deterministic Watson-Crick automaton M then the complementarity string of w must be in L and the restricted 1-limited deterministic Watson-Crick automaton must be in its final state after completely parsing w and its complementarity string in both its strands. Thus the non-deterministic pushdown automaton NP simulating M after parsing will have $ at the top of its stack and its current state will be of the form $(q_f,q_k)$ where $q_f \in F$, and $q_k \in F'$ and then the transition δ"$((q_f,q_k),\lambda,\$)=((q_f,q_k), \lambda)$ where $q_f \in F$, and $q_k \in F'$ can be applied to empty the stack and hence NP also accepts w.

If a string w is not accepted by the restricted 1-limited deterministic Watson-Crick automaton M. Then w is rejected due to occurrence of one of the following events.

1) The complementarity string of w is not in L, then the non-deterministic pushdown automaton NP simulating M at the end of parsing the input will never be in a state $(q_i,q_k)$ where $q_k \in F'$. Thus the transition $\delta''((q_f,q_k),\lambda,\$)=((q_f,q_k), \lambda)$ where $q_f \in F$, and $q_k \in F'$ cannot be applied and the stack is never empty hence NP also rejects w.

2) If the restricted 1-limited deterministic Watson-Crick automaton M after completely parsing both its strand is in a non-final state then the non-deterministic pushdown automaton NP simulating M at the end of parsing the input will never be in a state $(q_i,q_k)$ where $q_i \in F$. Thus the transition $\delta''((q_f,q_k),\lambda,\$)=((q_f,q_k), \lambda)$ where $q_f \in F$, and $q_k \in F'$ cannot be applied and the stack is never empty hence NP also rejects w.

3) The restricted 1-limited deterministic Watson-Crick automaton M halts before w is completely parsed. Two cases can arise in this condition.

a) The upper head is ahead of the lower head or vice-versa in either case the non-deterministic pushdown automaton NP simulating M will have a's or b's at the top of the stack thus the transition $\delta''((q_f,q_k),\lambda,\$)=((q_f,q_k), \lambda)$ where $q_f \in F$, and $q_k \in F'$ cannot be applied and the stack is never empty hence NP also rejects w.

b) The upper head is in the same position as the lower head in that case it may so happen that the transitions $\delta''((q_f,q_k),\lambda,\$)=((q_f,q_k), \lambda)$ where $q_f \in F$, and $q_k \in F'$ can be applied but even though the stack of NP is empty as NP simulates M it does not completely parse w thus NP rejects w.

So we can say that the non-deterministic pushdown automaton NP constructed from M and M' accepts the same language as the restricted 1-limited deterministic Watson-Crick automaton M.

**Corollary 3:** For every restricted deterministic Watson-Crick automaton where the restriction language is in unary regular languages there exists a non-deterministic pushdown automaton which accepts the same language as the restricted deterministic Watson-Crick automaton.

**Proof:** The proof follows from Lemma 1 and Theorem 7.

**Theorem 8:** The set of all languages accepted by restricted deterministic Watson-Crick automata where L is in unary regular languages is a proper subset of the context free languages.

**Proof:** From Corollary 3, we know that for every restricted deterministic Watson-Crick automaton where L is in unary regular languages there exists a non-deterministic pushdown automaton which accepts the same language. Thus the set of all languages accepted by restricted deterministic Watson-Crick automata where L is in unary regular languages is a subset of the context free languages. From [2] we know that there exists a context free language not accepted by any non-deterministic Watson-Crick automaton, hence there exists a context free language not accepted by any deterministic Watson-Crick automaton. We also know from Lemma 3, the set of all languages accepted by restricted deterministic Watson-Crick automata where L is regular is a subset of the set of all languages accepted by deterministic Watson-Crick automata as a result the set of all languages accepted by restricted deterministic Watson-Crick automata where L is in unary regular languages is also a subset of the set of all languages accepted by deterministic Watson-Crick automata. Therefore there exists a context free language which is not accepted by any restricted deterministic Watson-Crick automaton where L is in unary regular languages. Hence, the subset relation is proper.

**Theorem 9:** The set of all languages accepted by restricted deterministic Watson-Crick automata where L is in unary regular languages is a proper subset of the set of all languages accepted by deterministic Watson-Crick automata.

**Proof:** From Corollary 1, we know that for every restricted deterministic Watson-Crick automaton where L is regular there exists a deterministic Watson-Crick automaton which accepts the same language. Thus the set of all languages accepted by restricted deterministic Watson-Crick automata where L is in unary regular languages is a subset of the set of all languages accepted by deterministic Watson-Crick automata. From Theorem 8, we know that the set of all languages accepted by restricted deterministic Watson-Crick automata where L is in unary regular languages is a proper subset of the context free languages. From [3] we know that there exists a deterministic Watson-Crick automaton which accepts a context sensitive language. Hence, the subset relation is proper.

**Theorem 10:** If L is finite, the languages accepted by restricted deterministic Watson-Crick automata whose lower strand is restricted to L is also finite.

**Proof:** As L is finite there is only a finite number of strings in L which can act as complementarity strings to strings in $V^*$. Thus only those strings are allowed as input to the restricted deterministic Watson-Crick automata whose length matches to one of the strings in L and whose complementarity string is in L. As V is finite alphabet, number of such strings in $V^*$ whose length matches to one of the strings in L and whose complementarity string is in L is also finite. Hence a restricted deterministic Watson-Crick automaton whose lower strand is restricted to L will accept a subset of those strings in $V^*$ whose length matches to one of the strings in L and whose complementarity string is in L. As the subset of finite set is finite hence the language accepted by any restricted deterministic Watson-Crick automaton whose lower strand is restricted to L is finite.

**Theorem 11:** For every $L \subseteq V^*$, we can design a stateless restricted deterministic Watson-Crick automaton where lower strand is restricted to L and it has injective complementarity relation that accepts L.

**Proof:** The restricted deterministic Watson-Crick automaton $M=(V,\rho,Q,q_0,F,\delta,L)$ where lower strand is restricted to L which accepts L is defined as follows.

The finite alphabet is V, the set of states $Q=q_0$, the start state is $q_0$, the set of final states $F=\{q_0\}$, $\rho$ is the identity relation and $\delta$ has transitions of the form $q_0 \binom{x}{x} \to q_0$ and $x \in V$.

As the complementarity relation is identity the content of the lower strand and upper strand is exactly the same and as only those strings are allowed as input whose complementarity string is in L. Therefore the string in the upper strand of the restricted deterministic Watson-Crick automaton is also in L. The transition just ensures that both the head of the automaton reach the end of their respective strands so that the string in the upper strand can be accepted. Thus, this stateless restricted deterministic Watson-Crick automaton M where lower strand is restricted to L accepts L.

**Example 2:** The following restricted deterministic Watson-Crick automaton M is an example of a restricted deterministic Watson-Crick automaton whose lower strand is restricted to $L= \{a^{2n}b^n |n>1\}$ which accepts the language $L'=\{a^n b^n c^n | n>1\}$ where L' is a context sensitive language.

$M=(\{a,b,c\}, \{q_0, q_1, q_f\}, q_0, \{q_f\}, \delta, \rho, L)$, $L= \{a^{2n}b^n |n>1\}$ where the complementarity relation $\rho$ is as follows: $\rho(a)=a$, $\rho(b)=a$, $\rho(b)=c$ and the transitions in $\delta$ are as follows:

$q_0 \binom{a}{aa} \to q_1$, $q_0 \binom{b}{b} \to q_1$, $q_1 \binom{c}{\lambda} \to q_f$.

**Theorem 12:** There exists a restricted deterministic Watson-Crick automaton with L in context free languages which accepts a context sensitive language.

**Proof:** The proof follows from Example 2.

**Theorem 13:** The set of all languages accepted by deterministic Watson-Crick automata is a proper subset of the set of all languages accepted by restricted deterministic Watson-Crick automata with L in context free languages.

**Proof:** From Theorem 3, we know that the set of all languages accepted by deterministic Watson-Crick automata is equal to the set of all languages accepted by restricted deterministic Watson-Crick automata where L is regular and as all regular languages are context free languages thus, the set of all languages accepted by deterministic Watson-Crick automata is contained in the set of all languages accepted by restricted deterministic Watson-Crick automata with L in context free languages. Moreover if we consider L to be the palindrome language over $V^*$, which is a context free language by employing the restricted automaton described in Theorem 11, we can design a restricted deterministic automaton with L in context free languages that accepts the palindrome language. From [2] we know there is no non-deterministic Watson-Crick automaton that can accept the palindrome language hence there is no deterministic Watson-Crick automaton that can accept the palindrome language. Therefore, the subset relation is proper.

**Theorem 14:** Context sensitive languages is a subset of the set of all languages accepted by restricted deterministic Watson-Crick automata with L in context sensitive languages.

**Proof:** For every context sensitive language L' we can design a restricted deterministic Watson-Crick automaton with L in context sensitive languages by using the restricted automaton described in Theorem 11 and L'=L which accepts L'. Therefore context sensitive languages is a subset of the set of all languages accepted by restricted deterministic Watson-Crick automata with L in context sensitive languages.

**Theorem 15:** The set of all languages accepted by restricted deterministic Watson-Crick automata with L in context sensitive languages is a subset of context sensitive languages.

**Proof:** Here we state the intuition behind the proof. Given a restricted deterministic Watson-Crick automaton with L in context sensitive languages. We can think of simulating the restricted deterministic Watson-Crick automata using a non-deterministic linear bounded automaton whose working space is twice that of the input string. The non-deterministic linear bounded automaton guesses the complementarity string of the given input string non deterministically and then checks to see if the guessed complementary string belongs to the language L (where L is a context sensitive language). If yes, then it simulates the moves of the restricted deterministic Watson-Crick automaton on its tape and goes to an halting final state if the restricted deterministic Watson-Crick automaton reaches its final state at the end of traversing its input otherwise it halts in a non-final state. If no, then it rejects the input by halting in a non-final state.

Thus any restricted deterministic Watson-Crick automaton with L in context sensitive languages can be simulated by a non-deterministic linear bounded automaton. Thus the set of all languages accepted restricted deterministic Watson-Crick automata with L in context sensitive languages is a subset of context sensitive languages.

**Theorem 16:** The set of all languages accepted by restricted deterministic Watson-Crick automata with L in context sensitive languages and context sensitive languages are equal.

**Proof:** Proof follows from Theorem 14 and 15.

VI. CONCLUSION

In this paper, we introduce restricted deterministic Watson-Crick automaton where the string is the lower strand is restricted to a language L. We examine the computational power of the model with respect to L being in different language classes such as regular, unary regular, finite, context free and context sensitive. We show that restricted deterministic Watson-Crick automata with lower strand restricted to regular languages has the same computational power as deterministic Watson-Crick automata and set of all languages accepted by restricted deterministic Watson-Crick automata with lower strand restricted to unary regular languages is a proper subset of context free languages. We further show that set of all languages accepted by restricted deterministic Watson-Crick automata with lower strand restricted to context free languages is a proper super set of set of all languages accepted by deterministic Watson-Crick automata and that the set of all languages accepted by restricted deterministic Watson-Crick automata with lower strand restricted to context sensitive languages is equal to the context sensitive languages.

**REFERENCES**

[1] R.Freund, G.Paun, G.Rozenberg, A.Salomaa, Watson-Crick Finite Automata, Proc 3rd DIMACS Workshop on DNA Based Computers,Philadelphia, 297-328, 1997.

[2] G. Paun, G. Rozenberg, A. Salomaa, DNA Computing: New Computing Paradigms, Springer-Verlag, Berlin, 1998.

[3] E.Czeizler, E.Czeizler, L.Kari, K.Salomaa, On the descriptional complexity of Watson-Crick automata, Theoretical Computer Science, Volume 410, Issue 35, Pages 3250–3260, 28 August 2009.

[4] E. Czeizler, E. Czeizler, Parallel Communicating Watson-Crick Automata Systems, Proc. 11th International Conference AFL, 2005.

[5] E. Czeizler, E. Czeizler, On the Power of Parallel Communicating Watson-Crick Automata Systems, Theoretical Computer science (358)(1):142-147,2006.

[6]K. S Ray, K. Chatterjee, D. Ganguly, Equivalence of Subclasses of Two-Way Non-Deterministic Watson-Crick Automata, Applied Mathematics,Vol.4, No.10A, October 2013.

[7] E. Czeizler, E. Czeizler, A Short Survey on Watson-Crick Automata, Bull. EATCS 88 104-119, 2006.

[8] A.Paun, M.Paun,State and transition complexity of Watson-Crick finite automata,Fundamentals of Computation Theory,Lecture Notes in Computer Science, Volume 1684, pp 409-420,1999.

[9] K. S Ray, K. Chatterjee, D. Ganguly, State complexity of deterministic Watson–Crick automata and time varying Watson–Crick automata, Natural Computing, February 2015.